\documentclass[%
twocolumn,
superscriptaddress,
nofootinbib,
amsmath,amssymb,
aps,
prl,
]{revtex4-1}

\usepackage{graphicx}%
\usepackage{dcolumn}%
\usepackage{bm}%
\usepackage{dsfont}
\usepackage{mathtools}%

\usepackage[bookmarksnumbered,pdfpagelabels=true,plainpages=false,colorlinks=true, linkcolor=black,citecolor=black,urlcolor=black,breaklinks=true]{hyperref}

\newcommand{\prlsection}[1]{%
\textit{#1}---%
}

\usepackage[x11names,dvipsnames]{xcolor}
\usepackage{soul}
\usepackage{enumerate}

\begin{document}

\preprint{APS/123-QED}

\title{Effective action for relativistic hydrodynamics from Crooks fluctuation theorem}%

\author{Nicki Mullins}
\email{nickim2@illinois.edu}
\affiliation{Illinois Center for Advanced Studies of the Universe\\ Department of Physics, 
University of Illinois at Urbana-Champaign, Urbana, IL 61801, USA}

\author{Mauricio Hippert}%
\email{hippert.mauricio@ce.uerj.br}
\affiliation{Instituto de Física, Universidade do Estado do Rio de Janeiro, Rua São Francisco Xavier, 524, Rio de Janeiro, RJ, 20550-013, Brasil}

\author{Jorge Noronha}
\email{jn0508@illinois.edu}
\affiliation{Illinois Center for Advanced Studies of the Universe\\ Department of Physics, 
University of Illinois at Urbana-Champaign, Urbana, IL 61801, USA}

\date{\today}%

\begin{abstract}

A new effective theory framework for fluctuating hydrodynamics in the relativistic regime is derived using standard thermodynamical principles and general properties of non-equilibrium stochastic dynamics. 
For the first time,  we establish clear and concise conditions for ensuring that the resulting effective theories are causal, stable, and well-posed within general relativity.
These properties are independent of spacetime foliation and are valid in the full nonlinear regime. 
Out-of-equilibrium fluctuations are constrained by a relativistically covariant version of Crooks fluctuation theorem, 
which determines how the entropy production is distributed even 
when the system is driven by an external force. 
This leads to an emerging $\mathds{Z}_2$ symmetry responsible for imposing fluctuation-dissipation relations for n-point correlation functions, which matches the standard constraints for the Schwinger-Keldysh effective action.

\end{abstract}

\maketitle

\prlsection{Introduction}%
The relativistic expansion of the quark-gluon plasma observed in high-energy nucleus-nucleus collisions and its surprising fluid-like features \cite{Heinz:2013th} 
have motivated significant developments in the study of relativistic fluids~\cite{Romatschke:2017ejr, Rocha:2023ilf}. 
Most strikingly, these features seem to persist in smaller collision systems, such as proton-nucleus or proton-proton collisions, despite their significantly lower particle multiplicity \cite{Noronha:2024dtq}. 
The reduced size of these systems suggests that thermal fluctuations, which are usually neglected in state-of-the-art simulations, should grow in importance and be included in their hydrodynamic description \cite{SoaresRocha:2024afv}.   
Stochastic fluctuations are also expected to be relevant for the hydrodynamic behavior of the quark-gluon plasma near the putative QCD critical point \cite{Bzdak:2019pkr,An:2019csj}. 
Thus, a consistent formulation of relativistic fluctuating fluid dynamics is in order. 

A fully general-relativistic description of fluctuating hydrodynamics must preserve causality \cite{Choquet-Bruhat:2009xil}, in addition to satisfying 
the second law of thermodynamics and 
non-equilibrium fluctuation theorems \cite{Jarzynski_1997, 1998JSP....90.1481C, Crooks_1999, Harris_2007, Sevick_2008} in a covariant manner. 
Moreover, in the case of large deviations from equilibrium, such as those expected in heavy-ion collisions \cite{Gale:2012rq,Noronha-Hostler:2015coa}, or large fluctuations, such as those expected in the neighborhood of a second-order critical point \cite{landau2013statistical}, 
one should impose that these properties are satisfied in the fully nonlinear regime. 

In this Letter, we outline a new effective field theory (EFT) procedure for describing fluctuating relativistic fluids that uniquely meets the requirements above and displays several other new desired features that are not shared by any previous approach to this problem \cite{Kapusta:2011gt, Kovtun:2012rj, Haehl:2015pja, Haehl:2015uoc, Haehl:2016pec, Akamatsu:2016llw, Akamatsu:2017rdu, Mazeliauskas:2017wyz, An:2019osr, Kovtun:2014hpa, Harder:2015nxa, Sieberer:2015hba, Crossley:2015evo, Glorioso:2017fpd, Liu:2018kfw}. 
Our full EFT action can be specified by a single vector generating current and a dissipative potential related to entropy production. 
Causality and stability can be naturally encoded in the properties of this generating current, allowing for both to be imposed from the outset, unlike in other approaches.

When the system is causal and stable, it is also guaranteed that the corresponding action will have a positive-definite imaginary part, and the on-shell equations of motion will be both symmetric hyperbolic and flux-conservative. These properties ensure that the resulting dynamics are locally well-posed, i.e., 
given initial data, solutions
to the equations exist, are unique, and depend continuously on the data \cite{Kato1975TheCP, Choquet-Bruhat:2009xil, Disconzi:2023rtt}. This is mandatory when considering fluctuations around non-equilibrium solutions of the nonlinear equations of motion.

Nonlinear fluctuation-dissipation relations \cite{Wang:1998wg} are imposed through the Crooks fluctuation theorem \cite{1998JSP....90.1481C, Crooks_1999, Mallick_2011, Haehl:2015uoc, Torrieri:2020ezm}, derived here in covariant form.
In the absence of external fields, this procedure is similar to that developed in \cite{Huang:2023eyz, Guo:2022ixk}, and Crooks' theorem implies detailed balance. By coupling external sources and fields to conserved currents, we find that our effective action after evaluating the path integral satisfies a $\mathds{Z}_2$ symmetry, which is identical in form to the Kubo-Martin-Schwinger (KMS) symmetry 
in Schwinger-Keldysh theory \cite{Kubo:1957mj, Martin:1959jp, Kadanoff:1963axw, Sieberer:2015hba, Crossley:2015evo, Glorioso:2017fpd, Liu:2018kfw}. 
\emph{Notation}: We use $\hbar = k_B = c = 1$, a mostly plus metric $g_{\mu\nu}$, and anti-symmetrization as $A_{[\mu\nu]} = A_{\mu\nu}-A_{\nu\mu}$ .

\prlsection{Ideal hydrodynamics as a path integral}%
To set the stage, let us consider the ideal hydrodynamics of a system with an energy-momentum tensor $T^{\mu\nu}$ and current $J^\mu$ associated with a globally conserved $U(1)$ charge. We assume $T^{\mu\nu} = (\varepsilon+P)u^\mu u^\nu+g^{\mu\nu}P$ and $J^\mu = n u^\mu$, where $\varepsilon$ is the energy density, $P = P(\varepsilon,n)$ is the pressure defined by the equilibrium equation of state (EOS), $n$ is the number density, and $u^\mu$ is the timelike future-directed normalized fluid 4-velocity. The conservation laws, $\nabla_\mu T^{\mu\nu}=0$ and $\nabla_\mu J^\mu=0$, define the equations of motion for $n$, $\varepsilon$, and $u^\mu$. 

To formulate this as a path integral, we assume that the state of the system\footnote{In contrast to non-relativistic systems where the state is defined at a fixed time in an unambiguous manner, in relativistic systems the state of the system is defined across a suitable spacelike hypersurface \cite{Choquet-Bruhat:2009xil}.} can be described by  fields $\phi^a = (\alpha,\beta^\mu)$ together with a \emph{generating vector current} $X^\mu$ such that 
    $T^{\mu\nu} (\phi) \equiv {\partial X^{\mu}}/{\partial \beta_{\nu}}$
and    
    $J^{\mu}(\phi)  \equiv {\partial X^{\mu}}/{\partial \alpha}$. The initial and final states, $A$ and $B$, are specified by macroscopic fields $\phi_{(A)}(x\in \Sigma_A)$ and $\phi_{(B)}(x\in \Sigma_B)$, over the initial and final spacelike hypersurfaces $\Sigma_A$ and $\Sigma_B$, respectively. We want to recast the probability $ P[A \rightarrow B]$ of a macrostate $A$ going into a macrostate $B$ as a path integral over $\phi$, considering conservation laws as constraints. This can be done by introducing another set of auxiliary fields $\bar\phi$ 
\begin{equation} \label{Eq:transition_P}
    P[A \rightarrow B] =\int_A^B \mathcal{D} \phi\, \mathcal{D} \bar{\phi} \;
    e^{i \int d^4 x\, \sqrt{-g}\,\mathcal{L}(\phi, \bar{\phi})},
\end{equation}
where $\mathcal{L}= - \bar{\phi}^a \,\nabla_\mu\left({\partial X^{\mu}}/{\partial \phi^a}\right)$. Conservation laws can be recovered from this Lagrangian by variations with respect to $\bar\phi^a$. They emerge, in the context of Noether's theorem, from the symmetry of the action under shifts in $\bar\phi^a$. 

Ideal hydrodynamics is implemented by the assumption that $\beta^\mu$ is the only vector available so that the most general expression for the generating current is $X^\mu(\phi) = X(\alpha,\beta^2)\beta^\mu$. Defining $T\equiv (-\beta^2)^{-1/2}$ and $\mu \equiv \alpha \,T$, one can then use this $X^\mu$ to show that the computed $T^{\mu\nu}$ and $J^\mu$ take their ideal hydrodynamics form with    
$n \equiv (\partial X/\partial \mu)_T$, $u^\mu \equiv T\,\beta^\mu$,  
$\varepsilon\equiv n\,\mu + T\,s -X$, 
 and $s\equiv (\partial X/\partial T)_\mu$. Thus, the thermodynamic relations emerge from this construction with $X$ playing the role of the pressure $P$ of the system\footnote{We note the resemblance with divergence-type theories \cite{Geroch:1990bw, Calzetta:1997aj}.}. 
 Finally, we remark that since the path integral is obtained by integrating over solutions of the equations of motion, this construction is only possible if local well-posedness holds.

\prlsection{Dissipative hydrodynamics}%
Dissipative theories include nonconserved quantities that can be described in this formalism by adding out-of-equilibrium variables  $\varphi^a$ to the set of dynamic fields $\phi^a$, with their corresponding auxiliary variables $\bar\varphi^a$.    
The theory is specified by writing a generating current $X^\mu(\phi,\varphi)$,  
according to the symmetries under consideration. 
Because $\partial X^\mu/\partial \varphi^a$ should not be conserved, 
we add a \emph{dissipative potential} $\Xi(\phi^a,\varphi^a,\bar\varphi^a)$ to the Lagrangian 
\begin{equation}
\label{Eq:Lagrangian}
    \mathcal{L}= - \bar{\phi}^a \,\nabla_\mu\frac{\partial X^{\mu}}{\partial \phi^a} - \bar{\varphi}^a \,\nabla_\mu\frac{\partial X^{\mu}}{\partial \varphi^a}+ i\,\Xi(\phi^a,\varphi^a,\bar\varphi^a),
\end{equation}
which breaks the invariance of the action under shifts in $\bar\varphi^{a}$. 
This leads to the equations of motion
\begin{equation}
\label{Eq:EoM}
   \nabla_\mu \frac{\partial X^\mu}{\partial \phi^a}=0 \qquad \mathrm{and} \qquad -\nabla_\mu \frac{\partial X^\mu}{\partial \varphi^{a}} 
    + i\, \frac{\partial \Xi}{\partial \bar\varphi^a} = 0.
\end{equation}
Though only the first equation above encodes the conservation laws, both equations are still of conservative form.%
\footnote{Comparing to the notation of \cite{Geroch:1990bw}, the generating current $X^{\mu}$ is analogous to the vector $\chi^{\mu}$ while the dissipative potential is determined in their notation by $I_a = i (\partial \Xi / \partial \bar{\phi}^a)_{\bar{\Phi}=0}$. The structure of a divergence-type theory, as defined in \cite{Geroch:1990bw}, can be determined from $\chi^{\mu}, I_a$, but a given divergence-type theory does not necessarily correspond to a unique action.}
In general, the last term in Eq.~\eqref{Eq:Lagrangian}
can make the system non-deterministic, as nonlinear terms in $\bar\varphi^{a}$ will lead to equations of motion that are no longer constraints, but hold only on shell. This will be explored later in this paper.

\prlsection{Causality}%
To understand causality, it is useful to rewrite the equations of motion as $(\partial^2 X^{\mu} / \partial \Phi^a \partial \Phi^b) \nabla_{\mu} \Phi^b = i\,\partial \Xi/\partial\bar\Phi^a$, where $\Phi^a = (\phi^a,\varphi^a)$ denotes all the dynamical variables. The principal part of this equation of motion is the left-hand side (as long as $\Xi$ does not contain any derivatives), so causality comes from the properties of the vector-valued matrix $\partial^2 X^{\mu} / \partial \Phi^a \partial \Phi^b$, which is known as the characteristic matrix \cite{Choquet-Bruhat:2009xil}. Since $\partial^2 X^{\mu} / \partial \Phi^a \partial \Phi^b$ is a symmetric matrix, when $\partial^2 X^{0} / \partial \Phi^a \partial \Phi^b$ is positive-definite the system of partial differential equations is symmetric hyperbolic \cite{Choquet-Bruhat:2009xil}. Causality, in the full nonlinear regime, can be established directly by checking when $(\partial^2 X^{\mu} / \partial \Phi^a \partial \Phi^b) Z^a Z^b$ is timelike future-directed for any $Z^a \neq 0$ \cite{GEROCH1991394,Gavassino:2023odx}. Finally, we note that symmetric hyperbolic systems have a locally well-posed initial-value problem \cite{Kato1975TheCP,Choquet-Bruhat:2009xil}.

The conditions concerning causality and local well-posedness can be naturally implemented from the outset at the level of the generating current without any reference to the other parts of the action. This should be contrasted with other approaches, such as the Schwinger-Keldysh formalism, in which conditions for causality have to be determined on a case-by-case basis, and no systematic procedure within that approach is known that ensures causality in the nonlinear regime.

\prlsection{Covariant stability}%
We define the entropy current $s^\mu$ out of equilibrium in the non-ideal case as the Legendre transform of $X^\mu$~\cite{Geroch:1990bw}:
\begin{equation}
\label{Eq:entcurrent}
    s^\mu = X^\mu - \Phi^a \frac{\partial X^\mu}{\partial \Phi^a}.
\end{equation}
This leads to $\nabla_\mu s^\mu= -i \,\varphi^a\partial \Xi/\partial\bar\varphi^a$, which should be non-negative on shell in agreement with the second law since we consider an isolated fluid. 
As a matter of fact, the dissipative potential $\Xi$ is responsible for both dissipation and fluctuations and is constrained by the second law of thermodynamics. 
However, the fluctuation-dissipation theorem \cite{Callen:1951vq, Kubo:1957mj, Wang:1998wg} has not yet been enforced, so additional constraints are necessary. 
These further constraints can be found from Crooks fluctuation theorem \cite{1998JSP....90.1481C, Crooks_1999}, which we discuss in the next section.

The second law of thermodynamics states that the entropy of an isolated system always increases until equilibrium is reached. 
In covariant notation, we can write the total entropy in some spacelike hypersurface $\Sigma$ as $S[\Sigma] = \int_\Sigma d\Sigma_\mu\, s^\mu $, where $s^\mu$ is the entropy current. 
In relativity, the second law must hold for any spacelike foliation of spacetime $\Sigma(\tau)$, which implies the non-negativity of the entropy production rate $\nabla_\mu s^\mu \geq 0$.

In an isolated system, the thermodynamic equilibrium state corresponds to the state of maximum entropy taking into account the constraints from conserved quantities. 
Constraints on the total charge, energy, and momentum can be imposed by Lagrange multipliers $\alpha_*$ and $\beta_{*\nu}$, which couple to the conserved current $J^\mu$ and the energy-momentum tensor $T^{\mu\nu}$.  
Therefore, 
$\Omega[\Sigma]\equiv \int_\Sigma d\Sigma_\mu\, 
\Omega^\mu$
is minimized in equilibrium, with%
\footnote{The current $\Omega^\mu$ is closely related to the information current defined in \cite{Gavassino:2021kjm, Gavassino:2022roi}. In fact, both currents agree at quadratic order in deviations from equilibrium if we subtract from $\Omega^\mu$ its value at equilibrium.}
\begin{equation}
\label{Eq:Omega}
    \Omega^\mu \equiv  -s^\mu - \alpha_* J^\mu -\beta_{*\nu}\, T^{\mu\nu} = \delta \phi^a \frac{\partial X^{\mu}}{\partial \phi^a} + \varphi^a \frac{\partial X^\mu}{\partial \varphi^a} - X^{\mu} , 
\end{equation}
where $\delta \phi^a = \phi^a - \phi_*^a$ and $\phi_*\equiv (\alpha^*, \beta_*^\mu)$. We identify $\Omega[\Sigma]$ as the free energy, normalized by temperature, in the hypersurface $\Sigma$.
The constrained current $\Omega^\mu$ is stationary when $\delta \phi^a=0$, and $\varphi^{a} = 0$. Furthermore, since the Lagrange multipliers $\alpha_*$ and $\beta^\mu_*$ should only couple to the total charge, energy, and momentum, we have that $\nabla_\mu \alpha^*=0$ and $\nabla_\mu \beta^*_\nu + \nabla_\nu \beta^*_\mu=0$. 
Hence,  $\alpha=\alpha_*$, $\beta^\mu=\beta_*^\mu$, and $\varphi^{a} = 0$ defines the global equilibrium state.\footnote{The Killing condition for $\beta^*_\nu$ means that the background metric $g_{\mu\nu}$ used here must have a timelike Killing vector.}

Combined with the second law of thermodynamics, the conditions on $\alpha_*$ and $\beta^\mu_*$ imply that $\nabla_\mu \Omega^\mu = -\nabla_\mu s^\mu \leq 0$, so that $\Omega[\Sigma]$ can only decrease or stay the same, which makes it a Lyapunov functional for any spacelike $\Sigma$ if $\Omega^{\mu}$ is timelike future-directed. 
If we also enforce that $\Omega^{\mu}$ is timelike future-directed and $\delta\phi=\varphi^a=0$ is a global minimum of $n_{\mu} \Omega^{\mu}$ for any timelike past-directed $n^{\mu}$, it follows that the equilibrium state is stable for any spacelike foliation of spacetime, and hence for any inertial observer \cite{Gavassino:2021kjm}.

It is also possible to find novel connections between causality and stability that hold in the full nonlinear regime. Consider the 4-vector 
\begin{equation}
\label{Kcurrent}
    K^{\mu}(\Phi, \Phi') = \Omega^{\mu}(\Phi) - \Omega^{\mu}(\Phi') + (\Phi'^a - \Phi^a) \frac{\partial \Omega^{\mu}}{\partial \Phi^a} \Bigg|_{\Phi} .
\end{equation}
This is defined such that if $K^{\mu}$ is timelike future-directed for all $\Phi$ and $\Phi'$, then $n_{\mu} \Omega^{\mu}$ is a concave function for any timelike past-directed vector $n^{\mu}$. If the additional condition that $\Phi_*^a \equiv (\phi_*^a,\varphi^a=0)$ is a global minimum of $n_{\mu} \Omega^{\mu}$ is imposed, the equilibrium state system is stable. This is stronger than the standard Lyapunov stability conditions since the properties of $K^{\mu}(\Phi, \Phi')$ require concavity of $\Omega^{\mu}$ at all points in the space of dynamical variables, not just around the equilibrium state.%
\footnote{These conditions also ensure that $\Omega^{\mu}$ is stationary only in equilibrium.}
However, these conditions are useful because, if $K^{\mu}$ is timelike for some $\Phi, \Phi'$, then $(\partial^2 X^{\mu} / \partial \Phi^a \partial \Phi^b) \Phi^a \Phi'^b$ is also timelike. This provides a new connection between stability and causality that holds in the full nonlinear regime.%
\footnote{
A stronger notion of stability is necessary to draw this connection than in the linear case \cite{Gavassino:2021kjm}. For nonlinear systems, stability against a perturbation around the equilibrium state is not equivalent to stability against a perturbation around an arbitrary state. This is why $K^{\mu}(\Phi, \Phi')$ has two arguments, as it provides information about what happens when the system is in state $\Phi$ and then it is perturbed to state $\Phi'$.}

\prlsection{Crooks fluctuation theorem}%
So far, nothing guarantees that the dissipative dynamics described by Eq.~\eqref{Eq:Lagrangian} lead to the correct equilibrium probability distribution. %
In the absence of external fields, the principle of detailed balance ensures that the equilibrium distribution is a fixed point of the stochastic dynamics by imposing that 
the flow of probability due to any spontaneous fluctuation is balanced by the reverse process \cite{pitaevskii2012physical}. 
Mathematically, this involves the comparison of the transition probability $P[A \rightarrow B]$ to the reversed transition probability $P_{\bm\Theta}[\bm\Theta B \rightarrow \bm\Theta A]$, where $\bf\Theta$ is a discrete $\mathds Z_2$ transformation including time reversal.
Detailed balance has been used to study fluctuating relativistic hydrodynamics in the linear case \cite{Mullins:2023ott} and nonequilibrium dynamics of non-relativistic matter in \cite{Guo:2022ixk,Huang:2023eyz}.

A more general statement, valid in the presence of external parameters $\lambda^h(x)$, is given by the Crooks fluctuation theorem
\cite{1998JSP....90.1481C, Crooks_1999}, which holds even far from equilibrium. 
This theorem relates the probability of a ``trajectory" $\{\Phi(x),\lambda(x)\}$, given by $\mathcal{P}[\Phi|\lambda] =\int \mathcal{D} \bar{\Phi} \;
    e^{i \int d^4 x\, \sqrt{-g}\,\mathcal{L}(\Phi, \bar{\Phi}; \lambda)}$, to the probability of the reversed trajectory.
We enforce the aforementioned property by imposing that 
\begin{equation}
    \label{Eq:Crooks}
    \mathcal{P}[\Phi_a(x)\,|\,\lambda_h(x)] = \mathcal{P}_{\bm \Theta}[\Theta_a \,\Phi_a(-x) \,|\,\Theta_h \,\lambda_h(-x)] \;e^\omega, 
\end{equation}
where 
we define $\bm\Theta \Phi_a(x) \equiv \Theta_a \, \Phi_a(-x)$ and $\bm\Theta \lambda_h(x) \equiv \Theta_h \,\lambda_h(-x)$, with $\Theta_a,\, \Theta_h = \pm 1$ denoting the parity of the fields and external parameters%
\footnote{For instance, $\Theta_\varepsilon=1$, as the energy density $\varepsilon$ is even under parity and time reversal.}  and we do not sum over repeated indices.
Above, the total produced entropy in the spacetime region bounded by $\Sigma_A$ and $\Sigma_B$ is denoted by $\omega = \int d^4x\sqrt{-g} \,\sigma$. In terms of  the entropy production rate per unit volume $\sigma$,
Eq.~\eqref{Eq:Crooks} can be imposed by the condition
\begin{equation}
    \label{Eq:CrooksSym}
    \mathcal{L}_{\bm\Theta}(\bm\Theta\Phi,\bm\Theta\bar\Phi,\bm\Theta \lambda) = \mathcal{L}(\Phi,\bar\Phi,\lambda) - 
   i\, \sigma\,.
\end{equation}
Here, $\bm \Theta$ includes a parity transformation, and it also acts on the initial and final hypersurfaces $\Sigma_A$ and $\Sigma_B$.

Now we determine how the auxiliary fields $\bar\Phi$ transform under $\bm \Theta$. 
The entropy production should be odd and $X^\mu$ should be even under $\bm\Theta$, so that the first term in Eq.~\eqref{Eq:CrooksSym} can be recovered if $\bm \Theta \bar{\Phi}_a(x) = -\Theta_a \bar\Phi_a(-x)$. 
Next, we move to the transformation $\mathcal{L}\to \mathcal{L}_{\bm\Theta}$. 
This transformation is required because the microscopic time-reversal symmetry is broken by the change in the number of microstates in the stochastic process and its reverse. 
This is related to the second term in Eq.~\eqref{Eq:CrooksSym}, which must include the entropy production in the fluid, $\sigma_{\textrm{sys}}=\nabla_\mu s^\mu$, and can also include some external entropy production $\sigma_{\textrm{ext}}$ associated with the sources. 
To obtain that term, we follow the procedure in \cite{Mullins:2023ott} and take
$\mathcal{L}_{\bm\Theta}(\Phi,\bar\Phi,\lambda) = \mathcal{L}(\Phi,\bar\Phi - i \,\Phi,\lambda)$.
This is sufficient to satisfy Eq.~\eqref{Eq:CrooksSym} as long as $\Xi$ is left invariant, which leads to the following transformation rule for the reversal of a macroscopic process: 
\begin{subequations} \label{Eq:CrooksTransform}
\begin{align}
    &x^{\mu}  \rightarrow \bm\Theta x^{\mu} ,&  
    &\lambda^h  \rightarrow \bm\Theta \lambda^h,& \\
    &\Phi^a  \rightarrow \bm\Theta \Phi^a ,& 
    &\bar{\Phi}^a \rightarrow \bm\Theta \bar{\Phi}^a - i \bm\Theta \Phi^a, 
\end{align}
\vskip-20pt
\begin{align}
\label{Eq:CrooksXi}
 &\Xi \to \Xi,&  &\mathcal{L} \to \mathcal{L} 
+i \Phi^a \nabla_\mu\frac{\partial X^\mu}{\partial\Phi^a} 
 \,.&
\end{align}
\end{subequations}
In the absence of external work, the Lagrangian in \eqref{Eq:CrooksSym} changes only by a boundary term and the relations in \eqref{Eq:CrooksTransform} define the $\mathds{Z}_2$ symmetry associated with detailed balance discussed in \cite{Mullins:2023ott,Guo:2022ixk, Huang:2023eyz}. 

Comparing Eqs.~\eqref{Eq:CrooksSym} and \eqref{Eq:CrooksXi}, we find that $\sigma = - \Phi^a \,\nabla_\mu(\partial X^\mu/\partial\Phi^a)$. 
From Eq.~\eqref{Eq:entcurrent}, 
\begin{equation}
    \nabla_\mu s^\mu = -\nabla_\mu \Omega^{\mu} = \frac{\partial X^\mu}{\partial \lambda^h}\nabla_\mu \lambda^h - \Phi^a \nabla_\mu\frac{\partial X^\mu}{\partial\Phi^a} \,,
\end{equation}
so that we identify 
\begin{equation}
    \label{Eq:fullentprod}
    \sigma= \nabla_\mu s^\mu + \sigma_{\textrm{ext}} = - \Phi^a \,\nabla_\mu(\partial X^\mu/\partial\Phi^a)\,,
\end{equation}
where $\sigma_{\textrm{ext}}=(\partial X^\mu/\partial \lambda^h)\,\nabla_\mu \lambda^h$ 
is the entropy production in the reservoir responsible for setting the external parameters and fields $\lambda^h$.
The integrated entropy production is then $\omega = \Omega[B]-\Omega[A] +  \int d^4x \, \sqrt{-g} \, \sigma_{\textrm{ext}}$. 
Thus, we can identify $\sigma_{\textrm{ext}}$ as the work rate per unit volume realized by the system, normalized by its temperature, which connects our results to the standard form of the Crooks fluctuation  theorem \cite{1998JSP....90.1481C, Crooks_1999}.

\prlsection{Diffusion with an electric field}%
Let us now apply our approach to the case of charge diffusion in the presence of an Abelian gauge field $A_\mu$. 
The entropy produced by the reservoir responsible for fixing $A_\mu$ should be related to the work realized by the electric field, and we take
   $ \sigma_{\textrm{ext}}=(\partial X^\mu/\partial A^\nu)\,\nabla_\mu A_\nu =  \beta^\mu F_{\mu\nu} J^\nu$,
where $J^\mu = \partial X^\mu/\partial \alpha$ is the conserved current, and $F_{\mu\nu}\equiv \nabla_{[\mu} A_{\nu]}$ is the field strength tensor. 
Thus, we find that the gauge field couples to the generating current as $X^\mu = \ldots + \beta^{[\mu}J^{\nu]}A_\nu$, and to the Lagrangian as $\mathcal{L} = \ldots -  \bar\Phi^a\, \nabla_\mu [A_\nu\partial (\beta^{[\mu}J^{\nu]}) /\partial \Phi^a] $, where $\ldots$ denotes the terms that do not contain the gauge field.

The part of the Lagrangian that does not depend on the sources can be constructed by specifying a generating current and using the symmetry transformations in \eqref{Eq:CrooksTransform}. 
For simplicity, we take $\beta_{\mu}$ to be a constant timelike Killing vector, in which case a suitable generating current is $X^{\mu}(A_{\nu} = 0) = P(\alpha,j^2) \,\beta^{\mu} + \alpha j^{\mu}$. 
The conserved current is then $J^{\mu} = (\partial P / \partial \alpha) \beta^{\mu} + j^{\mu}$, with $j_\mu \beta^\mu=0$. Expanding the dissipative potential to leading order in the dynamical variables $\phi=\alpha$ and $\varphi^\mu = j^\mu$, we find that the Lagrangian for this theory takes the form 
\begin{equation}
\begin{split}
    \mathcal{L} = & -\bar{\alpha} \nabla_{\mu} J^{\mu} - \bar{j}_{\nu} \nabla_{\mu} \left( 2 \frac{\partial P}{\partial j^2} \beta^{\mu} \Delta^{\nu\lambda} j_{\lambda} + \alpha \Delta^{\mu\nu} \right) \\
    & + \frac{i}{\kappa} \Delta^{\mu\nu} \bar{j}_{\mu} \left( \bar{j}_{\nu} + i\,j_{\nu} \right) - \bar{j}_{\lambda} \nabla_{\mu} \beta^{[\mu} \Delta^{\nu]\lambda} A_{\nu} ,
\end{split}
\end{equation}
where $\kappa$ is an even function of the dynamical variables. From this, we recover the conservation law and a stochastic relaxation equation for the dissipative vector $j^{\mu}$. We note that the equations of motion are nonlinear. If we interpret $\partial P / \partial j^2$ as being proportional to $\tau / \kappa$ for some relaxation time $\tau$, the equation of motion for $j^{\mu}$ is analogous to the relaxation equations seen in Israel-Stewart transient hydrodynamics \cite{Israel:1979wp}. The steady state solution is defined by either $j^{\mu} = 0$ and $\nabla_{\mu} (\alpha \Delta^{\mu\nu}) + \beta_{\nu} F^{\mu\nu} = 0$, or by $\alpha=\alpha_*$ and $j^\mu = \kappa \,\beta_\nu F^{\mu\nu}$, depending on the boundary conditions. The first solution is the standard equilibrium behavior for a diffusive system in an external gauge field, while the latter defines Ohm's law.

\prlsection{Green's functions}%
To compute correlation functions, we need to introduce an extra source which couples to $J^\mu$ directly, which we denote as $\bar A_\mu$. 
This implies that in the presence of sources the Lagrangian becomes:
\begin{equation}
    \label{Eq:Lsource}
    \mathcal{L} = \ldots + \bar A_\mu J^\mu -  \bar\Phi^a\, \nabla_\mu \left[\frac{\partial \;}{\partial \Phi^a}(\beta^{[\mu}J^{\nu]})\, A_\nu\right]\,.
\end{equation}

Variations with respect to the gauge field $A_\mu$ and the source $\bar A_\mu$ allow us to compute any Green's function, as well as their higher-order generalizations  \cite{Wang:1998wg}. For instance, the retarded Green's function is
\begin{equation}
    \label{Eq:GR}
      G_R^{\mu\nu}(x,y) = \frac{\delta\langle J^\mu(x) \rangle}{\delta A_\mu(y)} = -\frac{\delta^2 W[A_\mu,\bar A_\mu]}{\delta A_\mu(y) \delta\bar A_\nu(x)}, 
\end{equation}
where $W$ is the effective action defined by 
\begin{equation}
    \label{Eq:EffAction}
    e^{i\,W[A_\mu,\bar A_\mu]}\equiv \int\mathcal{D} \Phi\, \mathcal{D} \bar{\Phi} \;
    e^{i \int d^4 x\, \sqrt{-g}\,\mathcal{L}(\Phi, \bar{\Phi}; A_\mu,\bar A_\nu)}.
\end{equation}
Every nonzero correlation function involves at least one derivative with respect to $\bar{A}_{\mu}$, so the standard Ward identities hold.

From Eqs.~\eqref{Eq:CrooksSym} and \eqref{Eq:Lsource}, we find that the effective action is symmetric under the usual KMS symmetry expected from effective actions on the Schwinger-Keldysh closed-time path contour in the classical limit \cite{Crossley:2015evo, Sieberer:2015hba, Glorioso:2017fpd, Liu:2018kfw}:
\begin{subequations} 
\label{Eq:KMS_transform}
\begin{align}
    \lambda^h & \rightarrow \mathbf{\Theta} \lambda^h , \\
    \bar \lambda^h & \rightarrow \mathbf \Theta \bar \lambda^h + i \mathbf \Theta \mathcal{L}_{\beta} \lambda^h ,
\end{align}
\end{subequations}
with the external sources $\lambda_a$ and $\bar \lambda_a$ set to $A_\mu$ and $\bar A_\mu$, respectively. 
The KMS symmetry in \eqref{Eq:KMS_transform} indicates that our effective action is compatible with $A_\mu$ and $\bar A_\mu$ being the symmetric and antisymmetric combinations of the gauge fields defined on each branch of the Schwinger-Keldysh contour, respectively. %

\prlsection{Fluctuation-dissipation theorem}%
The fluctuation-dissipation theorem can be recovered by considering the analytic structure of the path integral generating function. 
After evaluating the path integral in Eq.~\eqref{Eq:EffAction}, the generating function should have the form 
\begin{equation}
    W \sim \bar{\lambda}_a G_S^{ab} \bar{\lambda}_b + \lambda_a G_R^{ab} \bar{\lambda}_b + \bar{\lambda}_a G_A^{ab} \lambda_b + \ldots , 
\end{equation}
where $G_S^{ab}(x,x'), G_R^{ab}(x,x'), G_A^{ab}(x,x')$ are some set of correlation functions and $G_R^{ab} = - G_A^{ab \dagger}$. The $\ldots$ contains an infinite set of higher-point correlation functions. The generating function should be invariant under the symmetry in \eqref{Eq:KMS_transform}, which can be used to obtain relations between these correlation functions. Applying this transformation, we find that
\begin{equation}
\begin{split}
    W & \rightarrow \, \bar{\lambda}_a G_{S}^{ab} \bar{\lambda}_{b} + \left( i \mathcal{L}_{\beta} \lambda_a G_{S}^{ab} + \lambda_a G_{R}^{ab} \right) \bar{\lambda}_b \\
    & + \bar{\lambda}_{a} \left( G_{S}^{ab} i \mathcal{L}_{\beta} \lambda_b + G_{A}^{ab} \lambda_b \right)  + \lambda_a G_{R}^{ab} i \mathcal{L}_{\beta} \lambda_b \\
    & + i \mathcal{L}_{\beta} \lambda_a G_{S}^{ab} i \mathcal{L}_{\beta} \lambda_b + i \mathcal{L} \lambda_a G_{A}^{ab} \lambda_b .
\end{split}
\end{equation}
Upon Fourier transforming, we find that the generating function is only invariant if the correlation functions are related by 
\begin{equation}
    G_S^{ab}(\omega, k) = \frac{iT}{\omega} \left( G_R^{ab} - G_R^{ab\dagger} \right) = \frac{2T}{\omega} \mathrm{Im}(G_R^{ab}) . 
\end{equation}
If we identify $G_S^{ab}, G_R^{ab}, G_A^{ab}$ with the symmetrized, retarded, and advanced correlation functions respectively, this is the standard classical fluctuation-dissipation theorem \cite{Callen:1951vq, Kubo:1957mj}. This result is entirely expected since the path integral is invariant when the sources transform under the classical KMS symmetry, as in \cite{Crossley:2015evo, Glorioso:2017fpd, Liu:2018kfw}. This procedure can be generalized to higher-point correlation functions using standard arguments typically employed on the closed-time path \cite{Wang:1998wg}.

\prlsection{Building the action}
We summarize here the steps needed to build EFTs using our formalism\footnote{Detailed applications of our formalism will be presented in an upcoming companion paper.}:

\begin{enumerate}[(S.I)]
    \item \label{En:firststep}
    Choose the relevant equilibrium and out-of-equilibrium degrees of freedom $\Phi^a$. 
    \item \label{En:StepX}
    Define the generating current  $X^\mu(\Phi)$: start from the equilibrium equation of state and add the most general out-of-equilibrium corrections up to some maximum power in the dissipative fields $\varphi^{a}$, defining a systematic \emph{truncation scheme}.%
\footnote{Out-of-equilibrium corrections can then be systematically implemented via a power expansion in dissipative fields. This is similar to the so-called ``inverse Reynolds number" power counting of DNMR theory \cite{Denicol:2012cn}.}
    \item Couple external sources to the corresponding currents according to the transformation rules in \eqref{Eq:CrooksTransform}  or the symmetry in \eqref{Eq:KMS_transform} (up to boundary terms).
    \item \label{En:laststep}
    Construct the most general dissipative potential $\Xi$ that is invariant under the symmetry in \eqref{Eq:CrooksTransform} and contains all possible terms up to a maximum power of $\bar{\varphi}^{a}, \varphi^{a}$ that is equal to the power chosen in step (S.\ref{En:StepX}).
\end{enumerate}
In general, constraints on the different couplings are necessary to enforce physical evolution with the help of the following conditions: 
\begin{enumerate}[(C.I)]
    \item \label{En:firstcond}
    Impose that $K^\mu$ in \eqref{Kcurrent} is timelike future-directed to enforce causality, symmetric hyperbolicity, and stability of the equations of motion. 
    \item \label{En:lastcond}
    Use \eqref{Eq:fullentprod} to find the conditions that guarantee a non-negative total entropy production rate on shell. 
\end{enumerate}

\prlsection{Conclusions}%
In this Letter, we introduced a new systematic effective theory approach to relativistic hydrodynamics, and found conditions that lead to causal, symmetric hyperbolic, and stable evolution in the nonlinear regime, out of equilibrium, for any valid spacetime foliation. 
Stability for any foliation, 
together with the flux-conservative nature of the equations of motion, makes the resulting hydrodynamic theories well suited for numerical simulations, including through Monte Carlo techniques \cite{Florio:2021jlx, Schaefer:2022bfm, Florio:2023kmy, Basar:2024qxd, Chattopadhyay:2024jlh}. 
We also employed a covariant form of Crooks fluctuation theorem to constrain the stochastic dynamics in the presence of any prescribed spacetime dependent couplings or external fields, also 
finding the condition for positive on-shell entropy production in their presence.
This leads to couplings with external sources which display the dynamical KMS symmetry expected from EFT in the closed time path. 

Finally, we would like to comment on an important difference between our formalism and the Schwinger-Keldysh approach \cite{Harder:2015nxa, Crossley:2015evo, Sieberer:2015hba, Glorioso:2017fpd, Liu:2018kfw}. In the latter, both the fluctuation-dissipation theorem and the conservation laws also emerge from symmetry considerations. 
By truncating the constitutive relations to first-order in derivatives, the Schwinger-Keldish approach leads to an effective action for BDNK hydrodynamics \cite{Bemfica:2017wps, Bemfica:2019knx, Kovtun:2019hdm, Hoult:2020eho, Bemfica:2020zjp}. 
However, in the presence of noise, one can show that the imaginary part of the action is negative precisely in the regime where the hydrodynamic theory is causal and stable \cite{Mullins:2023ott, Jain:2023obu}, causing the stochastic path integral to be ill-defined in that case. In the linear regime, this issue can be circumvented by introducing additional UV regulators to the steady-state distribution \cite{Gavassino:2024ufs, Gavassino:2024vyu}. A proposal to address this issue via the introduction of extra fields in Schwinger-Keldish was worked out in \cite{Jain:2023obu}. However, in that work, no systematic procedure to remove the issues mentioned above and ensure causality and local well-posedness in a nonlinear regime was presented. Our formalism uniquely resolves these issues by creating a systematic framework that can be used to construct the first effective field theories for stochastic systems whose average evolution is manifestly causal and symmetric hyperbolic (hence locally well-posed). Furthermore, our approach can also be used to consider relativistic fluctuations around more general out-of-equilibrium steady states, as it is not limited to near-global equilibrium physics.

\section*{Acknowledgments}
N.M. and J.N. are partly supported by the U.S. Department of Energy, Office of Science, Office for Nuclear Physics under Award No. DE-SC0023861. M.H. was supported by Universidade Estadual do Rio de Janeiro, within the Programa de Apoio à Docência (PAPD).  

\bibliography{references}

\newpage
\appendix

\onecolumngrid
\section*{Supplemental Material}

\subsection*{A model for diffusion}

To illustrate how the effective theory procedure described in this work can be used to construct well-posed theories of relativistic fluctuating hydrodynamics, we return to the example of relativistic diffusion. Again, we take $\beta^{\mu}$ to be a constant timelike Killing vector so that the only dynamical conserved quantity is some conserved current $J^{\mu}$. Then, the relevant degrees of freedom are $\phi^a = \{ \alpha \}$ and $\varphi^a = \{ j^{\mu} \}$, and the generating current can be written as 
\begin{equation} \label{Eq:diffusive_X}
    X^{\mu} = P(\alpha, j^2) \beta^{\mu} + \alpha j^{\mu} + \beta^{[\mu} J^{\nu]} A_{\nu} ,
\end{equation}
for some external gauge field $A_{\mu}$. As before, the resulting conserved current is given by $J^{\mu} = (\partial P / \partial \alpha) \beta^{\mu} + j^{\mu}$, so we can interpret $j^{\mu}$ as a transverse, dissipative correction to the conserved current. Using the leading-order dissipative potential, the Lagrangian is again given by 
\begin{equation} \label{Eq:diffusive_L}
\begin{split}
    \mathcal{L} = & -\bar{\alpha} \nabla_{\mu} J^{\mu} - \bar{j}_{\nu} \nabla_{\mu} \left( 2 \frac{\partial P}{\partial j^2} \beta^{\mu} \Delta^{\nu\lambda} j_{\lambda} + \alpha \Delta^{\mu\nu} \right) + \frac{i}{\kappa} \Delta^{\mu\nu} \bar{j}_{\mu} \left( \bar{j}_{\nu} + i\,j_{\nu} \right) - \bar{j}_{\nu} \nabla_{\mu} \left( \beta^{[\mu} \Delta^{\lambda]\nu} A_{\lambda} \right) + \bar{A}_{\mu} J^{\mu} ,
\end{split}
\end{equation}
for some even function of the fields $\kappa$. The resulting on-shell equations of motion are given by 
\begin{subequations} \label{Eq:diffusive_EoMs}
\begin{align}
    \nabla_{\mu} \left( \frac{\partial P}{\partial \alpha} \beta^{\mu} + j^{\mu} \right) = \nabla_{\mu} J^{\mu} & = 0 , \\
    \nabla_{\mu} \left( 2 \frac{\partial P}{\partial j^2} \beta^{\mu} \Delta^{\nu\lambda} j_{\lambda} + \alpha \Delta^{\mu\nu} + \beta^{[\mu} \Delta^{\lambda]\nu} A_{\lambda} \right) & = \frac{1}{\kappa} j^{\nu} .
\end{align}
\end{subequations}
To compare this to standard hydrodynamic equations of motion, we define the relaxation time $\tau$ by $(\partial P / \partial j^2) = T \tau / 2 \kappa$, in which case the relaxation equation becomes 
\begin{equation}
    \nabla_{\mu} \left( \frac{\tau}{\kappa} u^{\mu} \Delta^{\nu\lambda} j_{\lambda} + \alpha \Delta^{\mu\nu} + \beta^{[\mu} \Delta^{\lambda]\nu} A_{\lambda} \right) = \frac{1}{\kappa} j^{\nu} .
\end{equation}
This now resembles a flux-conservative version of the Israel-Stewart relaxation equation \cite{Israel:1979wp}. Constructing this Lagrangian was rather straightforward; we simply posited the form of the generating current that provided the standard conserved current and everything else followed from symmetry. However, we have not yet verified that this is a well-posed theory by imposing physical constraints.

\subsubsection*{The second law of thermodynamics}

The simplest constraint to enforce is that of positive entropy production, as required by the second law of thermodynamics. Recall that the entropy current is given by $s^{\mu} = X^{\mu} - \Phi^a (\partial X^{\mu} / \partial \Phi^a)$. For the diffusive theory with generating current given by Eq.~\eqref{Eq:diffusive_X}, the resulting entropy current is given by 
\begin{equation} \label{Eq:diffusive_s}
    s^{\mu} = \left( P - \alpha \frac{\partial P}{\partial \alpha} - 2 j^2 \frac{\partial P}{\partial j^2} \right) \beta^{\mu} - \alpha j^{\mu} \, .
\end{equation}
Taking a derivative of this expression, using the equations of motion form Eq.~\eqref{Eq:diffusive_EoMs}, and adding the entropy production from external sources gives the total entropy production according to
\begin{equation}
    \sigma = \nabla_\mu s^\mu + \sigma_{\textrm{ext}} = - \Phi^a \,\nabla_\mu \frac{\partial X^\mu}{\partial\Phi^a} \,.
\end{equation}
Using this, we find that the total entropy production is given by 
\begin{equation}
    \sigma = \frac{1}{\kappa} j^2 \, .
\end{equation}
Since $j^{\mu}$ is spacelike, it follows that the entropy production is positive-definite as long as $\kappa$ is a positive-definite function of the fields. This is precisely equivalent to the stability condition that arises when we demand that the imaginary part of the action is positive-definite. To see this, we simply note that $\mathrm{Im} \mathcal{L} = \bar{j}^2 / \kappa$.

\subsubsection*{Correlation functions}

We can verify that the fluctuation-dissipation theorem holds by considering tree-level correlation functions. At linear order, the full Lagrangian can be written in the form 
\begin{equation} \label{Eq:Matrix_L}
    \mathcal{L} = -\frac{1}{2} \begin{pmatrix}
        \delta \Phi^a \\
        \delta \bar{\Phi}^a
    \end{pmatrix}^T \begin{pmatrix}
        0 & \frac{\partial^2 X_0^{\mu}}{\partial \Phi^a \partial \Phi^b} \nabla_{\mu} + \Xi_{ab} \\
        -\frac{\partial^2 X_0^{\mu}}{\partial \Phi^a \partial \Phi^b} \nabla_{\mu} + \Xi_{ab} & 2 i \Xi_{ab}
    \end{pmatrix} \begin{pmatrix}
        \delta \Phi^b \\
        \delta \bar{\Phi}^b
    \end{pmatrix} + \begin{pmatrix}
        \delta \Phi^a \\
        \delta \bar{\Phi}^a
    \end{pmatrix}^T \begin{pmatrix}
        \delta_{ab} & 0 \\
        0 & \delta_{ab} \mathcal{L}_{\beta}
    \end{pmatrix} \begin{pmatrix}
        \bar{\lambda}^b \\
        \lambda^b
    \end{pmatrix} ,
\end{equation}
where $\delta^{(2)} X^{\mu} = [(\chi T^2 / 2) \delta \alpha^2 + (\tau T / 2 \kappa) j^2]\beta^{\mu} + \delta \alpha j^{\mu}$ is the linearized generating current, and $\Xi_{AB}$ is given by 
\begin{equation}
    \Xi_{ab} = \begin{pmatrix}
        0 & 0 \\
        0 & \frac{1}{\kappa} \Delta_{\mu\nu}
    \end{pmatrix} .
\end{equation}
The resulting path integral is Gaussian, so it is possible to integrate out the dynamical variables $\Phi, \bar{\Phi}$. Assuming a Minkowski spacetime background, performing this integration, and Fourier transforming, we find that
\begin{equation}
    W[\lambda, \bar{\lambda}] = \frac{1}{2} \begin{pmatrix}
       \bar{\lambda}_{\alpha} \\
        \bar{\lambda}_j^{\mu} \\
        \frac{i\omega}{T}  \lambda_{\alpha} \\
        \frac{i\omega}{T}  \lambda_j^{\mu}
    \end{pmatrix}^T
    \begin{pmatrix}
        0 & 0 & -i \chi T \omega & i \Delta_{\lambda\mu} k^{\lambda} \\
        0 & 0 & i \Delta_{\lambda\nu} k^{\lambda} & \left( -i \frac{\tau}{\kappa} \omega - \frac{1}{\kappa} \right) \Delta_{\mu\nu} \\
        i \chi T \omega & -i \Delta_{\lambda\mu} k^{\lambda} & 0 & 0 \\
        -i \Delta_{\lambda\nu} k^{\lambda} & \left( i \frac{\tau}{\kappa} \omega - \frac{1}{\kappa} \right) \Delta_{\mu\nu} & 0 & \frac{2i}{\kappa} \Delta_{\mu\nu}
    \end{pmatrix}^{-1} 
    \begin{pmatrix}
        \bar{\lambda}_{\alpha} \\
        \bar{\lambda}_j^{\nu} \\
        -\frac{i\omega}{T}  \lambda_{\alpha} \\
        -\frac{i\omega}{T} \lambda_j^{\nu}
    \end{pmatrix} ,
\end{equation}
where $\omega = u^{\mu} k_{\mu}$. The standard retarded, advanced, and symmetrized correlation functions can then be obtained by taking variations with respect to suitable sources. Considering correlation functions in the chemical potential, we find that 
\begin{subequations}
\begin{align}
    G_R^{\alpha\alpha}(k) & = \frac{\kappa \mathbf{k}^2 / \chi}{T^2 \left( \kappa \mathbf{k}^2 - i \chi T \omega - \tau \chi T \omega^2 \right)} - \frac{1}{\chi T^2} , \\
    G_A^{\alpha\alpha}(k) & = \frac{\kappa \mathbf{k}^2 / \chi}{T^2 \left( \kappa \mathbf{k}^2 + i \chi T \omega - \tau \chi T \omega^2 \right)} + \frac{1}{\chi T^2} , \\
    G_S^{\alpha\alpha}(k) & = \frac{2 \kappa \mathbf{k}^2}{|\kappa \mathbf{k}^2 - i \chi T \omega - \tau \chi T \omega^2|^2} . 
\end{align}
\end{subequations}
This symmetrized self-correlation function (as well as $G_S^{j_{\mu}j_{\nu}}$) is positive-definite as long as $\kappa, T > 0$, which is precisely the requirement we found from the second law of thermodynamics. Up to contact terms, these are the standard correlation functions for Israel-Stewart diffusion obtained in \cite{Mullins:2023ott}. These correlation functions are related by $G_S^{\alpha\alpha} = (i T /\omega) (G_R^{\alpha\alpha} - G_A^{\alpha\alpha})$, which is the standard fluctuation-dissipation relation. As expected, the KMS symmetry ensures that correlation functions obey the expected relations. Nonlinear correlations can then be understood using perturbative techniques. 

In general, the form of the symmetrized correlation functions can be determined from Eq.~\eqref{Eq:Matrix_L}. We find that the symmetrized correlation functions are given by 
\begin{equation} \label{Eq:GS_gen}
    G_S^{ab} = \left( -i \frac{\partial^2 X_0^{\mu}}{\partial \Phi^a \partial \Phi^c} k_{\mu} + \Xi_{ac} \right)^{-1} \Xi^{-1}_{cd} \left( i \frac{\partial^2 X_0^{\mu}}{\partial \Phi^d \partial \Phi^b} k_{\mu} + \Xi_{db} \right)^{-1} ,
\end{equation}
where $\Xi^{-1}_{cd}$ denotes the inverse of $\Xi_{cd}$ within the dissipative subspace. Recalling that the entropy production is given by $\varphi^a \Xi_{ab} \varphi^b$, the second law of thermodynamics requires that $\Xi_{ab}$ is positive-definite within the dissipative subspace. Note that $\Xi_{ab}$ should be positive-definite in the dissipative subspace, not positive semi-definite so that all the dissipative degrees of freedom contribute to entropy production. Since the inverse of a positive-definite matrix is also positive-definite, the self-symmetrized correlation functions given by Eq.~\eqref{Eq:GS_gen} are always positive-definite when the second law of thermodynamics holds.

\subsubsection*{Causality, stability, and symmetric hyperbolicity}

Causality can be understood by considering the properties of the characteristic matrix, $\partial^2 X^{\mu} / \partial \Phi^a \partial \Phi^b$. For this example of diffusion, the characteristic matrix is given by 
\begin{equation}
    \frac{\partial^2 X^{\mu}}{\partial \Phi^a \partial \Phi^b} = \begin{pmatrix}
        \frac{\partial^2 P}{\partial \alpha^2} \beta^{\mu} & 2 \frac{\partial^2 P}{\partial j^2 \partial \alpha} \beta^{\mu} j^{\nu} + \Delta^{\mu\nu} \\
        2 \frac{\partial^2 P}{\partial \alpha \partial j^2} \beta^{\mu} j^{\lambda} + \Delta^{\mu\lambda} & 4 \frac{\partial^2 P}{\partial (j^2)^2} \beta^{\mu} j^{\nu} j^{\lambda} + 2 \frac{\partial P}{\partial j^2} \beta^{\mu} \Delta^{\nu\lambda}
    \end{pmatrix} .
\end{equation}
Given that the partial derivatives with respect to the dynamical fields commute, this matrix is symmetric. The equations of motion are symmetric hyperbolic in a given frame if $\partial^2 X^0 / \partial \Phi^a \partial \Phi^b$ is positive-definite.
We will start by considering the hyperbolicity conditions in the local rest frame, in which case $\beta^\mu = (1/T,0,0,0)$ and $j^\mu = (0,j^i)$, and
\begin{equation}
    \frac{\partial^2 X^{0}}{\partial \Phi^a \partial \Phi^b} = \frac{1}{T} \begin{pmatrix}
        \frac{\partial^2 P}{\partial \alpha^2} & 2 \frac{\partial^2 P}{\partial j^2 \partial \alpha} j^{\nu} \\
        2 \frac{\partial^2 P}{\partial \alpha \partial j^2} j^{\lambda} & 4 \frac{\partial^2 P}{\partial (j^2)^2}j^{\nu} j^{\lambda} + 2 \frac{\partial P}{\partial j^2} \Delta^{\nu\lambda}
    \end{pmatrix} .
\end{equation}
Hyperbolicity in the local rest frame is enough to determine hyperbolicity in a general frame if the theory is causal, which we will show below.
This matrix is positive-definite if all of its leading principal minors are positive. We therefore consider the series of determinants
\begin{subequations} \label{Eq:determinants_gen}
\begin{align}
    \mathrm{det} \frac{\partial^2 X^{0}}{\partial \Phi^a \partial \Phi^b} & \propto \frac{4}{T^3} \frac{\partial P}{\partial j^2} \left[ \frac{\partial^2 P}{\partial \alpha^2} \left( \frac{\partial P}{\partial j^2} + 2 j^2 \frac{\partial^2 P}{\partial (j^2)^2} \right) - 2 j^2 \left( \frac{\partial^2 P}{\partial \alpha \partial j^2} \right)^2 \right] , \\
    \mathrm{det} \left( 4 \frac{\partial^2 P}{\partial (j^2)^2}j^{\nu} j^{\lambda} + 2 \frac{\partial P}{\partial j^2} \Delta^{\nu\lambda} \right) & \propto \frac{4}{T^2} \frac{\partial P}{\partial j^2} \left( \frac{\partial P}{\partial j^2} + 2 j^2 \frac{\partial^2 P}{\partial (j^2)^2} \right) , \\
    \mathrm{det} \Delta_{(j)\nu}^{\rho} \left( 4 \frac{\partial^2 P}{\partial (j^2)^2}j^{\nu} j^{\lambda} + 2 \frac{\partial P}{\partial j^2} \Delta^{\nu\lambda} \right) & \propto \frac{2}{T} \frac{\partial P}{\partial j^2} ,
\end{align}
\end{subequations}
where $\Delta_{(j)\nu}^{\rho}$ is the projector orthogonal to both $\beta^{\mu}$ and $j^{\mu}$. After this projection, the matrix is proportional to the identity, so no further sub-matrices need be considered. Note that the proportionality constant in each case involves a number of factors of $(\partial P / \partial j^2)$ that depend on the number of spatial dimensions. From these three determinants, we can obtain the hyperbolicity conditions in the local rest frame
\begin{subequations}
\begin{align}
    \frac{\partial^2 P}{\partial \alpha^2} & > 0 , \\
    \frac{\partial P}{\partial j^2} & > 0 , \\
    \frac{\partial^2 P}{\partial \alpha^2} \frac{\partial^2 P}{\partial (j^2)^2} & \geq \left( \frac{\partial^2 P}{\partial \alpha \partial j^2} \right)^2 .
\end{align}
\end{subequations}
The first condition is the standard concavity of the equilibrium equation of state, the second enforces that the relaxation time is positive-definite, and the final condition implies that $(\partial^2 P / \partial (j^2)^2) > 0$ which is another concavity condition.

This can be understood further by performing an inverse-Reynolds expansion of the pressure, in which case we define $P = P_0(\alpha) + \sum_{n=1}^N P_n(\alpha) j^{2n} / 2n$ for some maximum power $N$. At leading order $N=1$, one finds 
\begin{subequations}
\begin{align}
    \mathrm{det} \frac{\partial^2 X^{0}}{\partial \Phi^a \partial \Phi^b} & \propto \frac{P_1}{2 T^3} \left[ 2 P_1 P_0'' + j^2 \left( P_1 P_1'' - 2 (P_1')^2 \right) \right] , \\
    \mathrm{det} \left( 4 \frac{\partial^2 P}{\partial (j^2)^2} j^{\nu} j^{\lambda} + 2 \frac{\partial P}{\partial j^2} \Delta^{\nu\lambda} \right) & \propto \left( \frac{2 P_1}{T} \right)^2 , \\
    \mathrm{det} \Delta_{(j)\nu}^{\rho} \left( 4 \frac{\partial^2 P}{\partial (j^2)^2}j^{\nu} j^{\lambda} + 2 \frac{\partial P}{\partial j^2} \Delta^{\nu\lambda} \right) & \propto \frac{2 P_1}{T} ,
\end{align}
\end{subequations}
where the prime denotes differentiation with respect to $\alpha$. If the term in parentheses becomes negative, then we can always choose a value of $j^2$ such that the determinant becomes zero. So, the determinant is only nonzero for all values of $\alpha, j^2$ if $P_1 P_1'' \geq 2 (P_1')^2$. 

From here, it is useful to take a brief detour to consider the stability of the equilibrium state. This can be considered by deriving the conditions for the perturbation of $\Omega^{\mu} = \delta \Phi^a (\partial X^{\mu} / \partial \Phi^a) - X^{\mu}$ around equilibrium to be a Lyapunov functional. All the conditions discussed in \cite{Gavassino:2021kjm} hold by construction, except the condition that the perturbation of $\Omega^{\mu}$ around equilibrium should be timelike future-directed. For the leading order theory, we have
\begin{equation}
    \Omega^{\mu} = \left[ \left( \delta \alpha \frac{\partial P_0}{\partial \alpha} - P_0 \right) + \frac{1}{2} \left( \delta \alpha \frac{\partial P_1}{\partial \alpha} + P_1 \right) j^2 \right] \beta^{\mu} + \delta \alpha j^{\mu} .
\end{equation}
The conditions for perturbations of this vector to be timelike future-directed can be determined by contracting with a timelike past-directed vector $n^{\mu} = (-1, \mathbf{n})$. A vector $V^{\mu}$ is then timelike future-directed if $n_{\mu} V^{\mu} > 0$ for all $\mathbf{n}$ with $\mathbf{n}^2 < 1$, so we compute
\begin{equation}
\begin{split}
    n_{\mu} (\Omega^{\mu} - \Omega_0^{\mu}) & = \frac{1}{T} \left[ \delta \alpha \frac{\partial P_0}{\partial \alpha} - \delta P_0 + \frac{1}{2} \left( \delta \alpha \frac{\partial P_1}{\partial \alpha} + P_1 \right) j^2 \right] + \delta \alpha n_i j^i , \\
    & = \frac{1}{2T} \left( \delta \alpha \frac{\partial P_1}{\partial \alpha} + P_1 \right) \left[ \frac{T \delta \alpha}{\delta \alpha \frac{\partial P_1}{\partial \alpha} + P_1} n^i + j^i \right]^{2} + \frac{\delta \alpha}{T} \frac{\partial P_0}{\partial \alpha} - \frac{1}{T} \delta P_0 - \frac{T n^2 \delta \alpha^2}{2 \left( \delta \alpha \frac{\partial P_1}{\partial \alpha} + P_1 \right)} .
\end{split}
\end{equation} 
Here, $\delta P_0$ is given by $\delta P_0 = P_0 - P_0(\alpha_0)$. It follows that the system is stable if 
\begin{subequations} \label{Eq:thermo_stability_conds}
\begin{align}
    \delta \alpha \frac{\partial P_1}{\partial \alpha} + P_1 & > 0 , \\
    \left( \delta \alpha \frac{\partial P_0}{\partial \alpha} - \delta P_0 \right) \left( \delta \alpha \frac{\partial P_1}{\partial \alpha} + P_1 \right) & \geq \frac{T^2}{2} \delta \alpha^2 . 
\end{align}
\end{subequations}
The first condition can only be simultaneously satisfied with the earlier hyperbolicity condition that $P_1 P_1'' \geq 2 (P_1')^2$ if $P_1$ is independent of $\alpha$. Then, the hyperbolicity conditions reduce to $P_0'' > 0$ and $P_1 > 0$. In general, hyperbolicity and stability together will require that $P_N$ is independent of $\alpha$, for any maximum power in the inverse-Reynolds expansion $N$. 

Symmetric hyperbolicity on its own is not sufficient to ensure a sensible relativistic initial value problem, as we must also find the conditions for causality. These follow from the requirement that $(\partial^2 X^{\mu} / \partial \Phi^a \partial \Phi^b) Z^a Z^b$ is a timelike future-directed vector for any $Z^a \neq 0$. Again, this can be shown by contracting the characteristic matrix with a timelike past-directed vector in the form $n^{\mu} = (-1, \mathbf{n})$ with $\mathbf{n}^2 < 1$. 
Then, causality follows from the condition
\begin{equation}
    n_{\mu} \frac{\partial^2 X^{\mu}}{\partial \Phi^a \partial \Phi^b} Z^a Z^b > 0 .
\end{equation}
This is equivalent to the statement that $n_{\mu} (\partial^2 X^{\mu} / \partial \Phi^a \partial \Phi^b)$ is a positive-definite matrix; here it will be easier to prove this by showing that it has positive-definite eigenvalues. Again working at leading order in the inverse-Reynolds expansion and taking the local rest frame of the fluid, we find that 
\begin{equation}
    n_{\mu} \frac{\partial^2 X^{\mu}}{\partial \Phi^a \partial \Phi^b} = \begin{pmatrix}
        \frac{1}{T} P_0'' & n^i \\
        n^j & \frac{P_1}{T} \delta^{ij}
    \end{pmatrix} ,
\end{equation}
where we have assumed that $P_1$ is a constant in accordance with the stability and hyperbolicity conditions. The unique eigenvalues of this matrix are given by 
\begin{subequations}
\begin{align}
    E_0 & = \frac{P_1}{T} , \\
    E_{\pm} & = \frac{1}{2T} \left( P_0'' + P_1 \pm \sqrt{(P_0'' - P_1)^2 + 4 T^2 \mathbf{n}^2} \right) .
\end{align}
\end{subequations}
The first eigenvalue $E_0$ is positive-definite when $P_1 > 0$, so we again find that the relaxation time must be positive. The eigenvalue $E_+$ is positive-definite when $P_0'', P_1 > 0$, so we again find the convexity condition for the equation of state. Finally, $E_-$ is minimized when $\mathbf{n}^2 \rightarrow 1$, in which case we obtain the condition $P_0'' + P_1 > \sqrt{(P_0'' - P_1)^2 + 4T^2}$, which implies that
\begin{equation}
    P_1 P_0'' > T^2 . 
\end{equation}
This is the only causality condition that does not follow from hyperbolicity; it can be thought of as bounding the minimum relaxation time for a given equation of state.

\end{document}